\documentclass[journal]{IEEEtran}
\IEEEoverridecommandlockouts
\usepackage{cite}
\usepackage{amsmath,amssymb,amsfonts,mathtools}
\usepackage{algorithmic}
\usepackage{graphicx}
\usepackage{textcomp}
\usepackage{tikz}
 \usepackage{pgfplots}
 \usepackage{pgfplotstable}
\usepackage{subfigure}
\usepackage{multirow}
\newcommand{\foo}{\foo}
\usetikzlibrary{shadows}
\usetikzlibrary{patterns}
\usetikzlibrary{positioning}
\usepgfplotslibrary{fillbetween}
\usetikzlibrary{patterns}
\usetikzlibrary{calc}

\newtheorem{exampleplain}{Example}

\usetikzlibrary{arrows,arrows.meta,positioning,shadows,shapes,backgrounds,calc,fit}

\def\BibTeX{{\rm B\kern-.05em{\sc i\kern-.025em b}\kern-.08em
    T\kern-.1667em\lower.7ex\hbox{E}\kern-.125emX}}

\pgfmathdeclarefunction{gauss}{3}{%
  \pgfmathparse{1/(#3*sqrt(2*pi))*exp(-((#1-#2)^2)/(2*#3^2))}%
}

\begin{document}

\title{A Soft-Aided Staircase Decoder Using Three-Level Channel Reliabilities\\
}

\author{Yi~Lei, Bin~Chen,~\IEEEmembership{Member,~IEEE}, Gabriele~Liga,~\IEEEmembership{Member,~IEEE}, Alexios
Balatsoukas-Stimming,~\IEEEmembership{Member,~IEEE}, Kaixuan~Sun, Alex~Alvarado,~\IEEEmembership{Senior~Member,~IEEE}
\thanks{This work is partially supported by the NSFC Program (No. 62001151) and Fundamental Research Funds for the Central Universities (JZ2020HGTA0072, JZ2020HGTB0015), and Anhui Provincial Natural Science Foundation (2008085QF282).
The work of G. Liga is funded by the EUROTECH postdoc programme under the European Union’s Horizon 2020 research and innovation programme (Marie Skłodowska-Curie grant agreement No 754462). The work of A. Alvarado is supported by the Netherlands Organisation for Scientific Research (NWO) via the VIDI Grant ICONIC (project number 15685) and the European Research Council (ERC) under the European Unions Horizon 2020 research and innovation programme (grant agreement No. 757791).
\emph{(Corresponding author: bin.chen@hfut.edu.cn)}}
\thanks{Y. Lei and B. Chen are with the School of Computer Science and Information Engineering, Hefei University of Technology, China (\mbox{e-mails:} \{leiyi, bin.chen\}@hfut.edu.cn).}

\thanks{A.
Balatsoukas-Stimming is with the Electronic Systems group, Department of Electrical Engineering, Eindhoven University of Technology, The Netherlands (\mbox{e-mail:} a.k.balatsoukas.stimming@tue.nl).}

\thanks{K. Sun is with State Key Laboratory of Information of Photonics and Optical Communications, Beijing University of Posts and Telecommunications, China (\mbox{e-mail: sunkaixuan@bupt.edu.cn}).}

\thanks{G. Liga and A. Alvarado are with the Signal Processing Systems (SPS) Group, Department of Electrical Engineering, Eindhoven University of Technology, The Netherlands (\mbox{e-mails:} \{g.liga, a.alvarado\}@tue.nl).}
}

\maketitle

\begin{abstract}
The soft-aided bit-marking (SABM) algorithm is based on the idea of marking bits as highly reliable bits (HRBs), highly unreliable bits (HUBs), and uncertain bits to improve the performance of hard-decision (HD) decoders. The HRBs and HUBs are used to assist the HD decoders to prevent miscorrections and to decode those originally uncorrectable cases via bit flipping (BF), respectively. In this paper, an improved SABM algorithm (called iSABM) is proposed for staircase codes (SCCs). Similar to the SABM, iSABM marks bits with the help of channel reliabilities, i.e., using the absolute values of the log-likelihood ratios. The
improvements offered by iSABM include: (i) HUBs being classified using a reliability threshold, (ii) BF randomly selecting HUBs, and (iii) soft-aided decoding over multiple SCC blocks. The decoding complexity of iSABM is comparable of that of SABM. This is due to the fact that on the one hand no sorting is required (lower complexity) because of the use of a threshold for HUBs, while on the other hand multiple SCC blocks use soft information (higher complexity).
Additional gains of up to $0.53$~dB with respect to SABM and $0.91$~dB with respect to standard SCC decoding at a bit error rate of $10^{-6}$ are reported. Furthermore, it is shown that using $1$-bit reliability marking, i.e., only having HRBs and HUBs, only causes a gain penalty of up to $0.25$~dB with a significantly reduced memory requirement.
\end{abstract}


\section{Introduction}

Forward error correction (FEC) is an essential ingredient for achieving reliable data transmission in modern optical communication systems. FEC decoders typically come in two flavors: soft-decision (SD) and hard decision (HD). SD-FEC decoders are typically used for example to decode low density parity check codes (LDPC) and provide large coding gains. However, SD-FEC decoders pose implementation challenges in terms of complexity, delay, power consumption and circuit area~\cite{Pillai_JLT2014}. As targeted data rates exceed $400$~Gbps, simple but powerful HD-FEC decoders are more attractive for future high-speed low-cost optical transport networks (OTNs).

Staircase codes (SCCs)~\cite{Smith2012}, which use HD decoders in an iterative fashion, have become particularly interesting to OTNs in recent years. SCCs are built on simple component codes, e.g., Bose-Chaudhuri-Hocquenghem (BCH) codes, and are iteratively decoded by bounded-distance decoding (BDD). SCCs currently have been recommended for $100$G long-reach (LR) OTNs~\cite{G709.2}, flexible LR OTNs~\cite{G709.3}, and $400$G extended long-reach (ZR) OTNs (as an outer code)~\cite{OIF400G}. However, limited by the HD nature of the decoder, standard SCCs give significant performance losses when compared to SD decoding. For this reason, the design of SCC decoding algorithms with higher coding gains and low decoding complexity has attracted much attention in recent years.

Early works on improved SCC decoding took advantage of the staircase structure of SCCs~\cite{SmithPhD,Christian1, Holzbaur2017}. Since each bit in the staircase structure is protected by two component codewords, one can identify (and prevent) miscorrections by checking conflicts between the two component codewords~\cite{SmithPhD,Christian1}, or locate (and solve via bit flipping) stall patterns through the intersections of nonzero-syndrome component words~\cite{Holzbaur2017}. Although these methods are simple as they only operate on binary messages, their gains are limited. To obtain a higher gain, an extreme solution is to completely replace the BDD component decoder with a SD decoder. This was proposed in~\cite{Douxin_ISTC2018,ZhouWCSP2018SCCPolar,SCC_LDPC2020,CondoOFC2020}, where channel soft information, i.e., log-likelihood ratios (LLRs), were used. However, this solution has a greatly increased decoding complexity.

A new class of decoding schemes, called soft-aided HD (SA-HD) decoders, have been shown to provide a good compromise between complexity and performance. The main principle of SA-HD decoders is to assist the HD decoding with channel LLRs, while keeping the message exchange between the component decoders binary. For example,~\cite{Alireza,AlirezaSCC} proposed to make a hard decision based on the weighted sum of the BDD output and the channel LLR, while~\cite{AlirezaarXiv2019} replaced the BDD component decoder with generalized minimum distance decoding, which introduces erasures according to the channel LLRs. The work in~\cite{Alireza2020} is an enhanced version of~\cite{Alireza}, which improves the combining rule by deriving a more accurate estimate of the reliability of the BDD outputs. Lately, a so-called BEE-SCC algorithm was proposed in~\cite{AlirezaSCC2020}. BEE-SCC extends the work in~\cite{Alireza2020} by using an extra decoding attempt based on error and erasure decoding of the component codes. The results in~\cite{AlirezaSCC2020} show that BEE-SCC can achieve gains up to $0.88$ dB with respect to standard SCCs. Due to the high coding gain and low complexity, SA-HD decoders are thought to be a promising decoding scheme towards the future high-throughput optical fiber communications~\cite{AlexandreArXiv2019}.

Recently, we have proposed a SA-HD decoder based on a soft-aided bit-marking (SABM) algorithm to improve the performance of SCCs~\cite{YiISTC2018,YiTCOM2019}. We will refer to this algorithm as SABM-SCC decoder. Different from the methods proposed in~\cite{Alireza,AlirezaSCC,AlirezaarXiv2019,Alireza2020,AlirezaSCC2020}, the SABM-SCC decoder only uses soft information to mark bits as highly reliable bits (HRBs) and highly unreliable bits (HUBs). Using HRBs, the SABM-SCC decoder prevents miscorrections by checking whether the flipped bits are in conflict with HRBs. Via the HUBs, the SABM-SCC decoder flips a certain number of HUBs, which are the most likely errors, to handle BDD failures and miscorrections. The results in~\cite{YiISTC2018,YiTCOM2019} show that the SABM-SCC decoder can yield up to $0.30$~dB additional gain compared to standard SCC decoder at a bit-error ratio (BER) of $10^{-7}$. This additional gain has been experimentally demonstrated to provide $240$~km reach extension in a coherent optical fiber communication system~\cite{BinOFC2019}.

The SABM algorithm with minor modifications was also recently demonstrated to perform well for another popular HD-FEC scheme, i.e., product codes (PCs)~\cite{YiTCOM2019}. Improvements of up to $0.5$~dB were achieved with respect to standard decoding of PCs. The latest work in~\cite{Gabriele_ECOC2019} shows that SABM with scaled reliability (SABM-SR) can improve the coding gains up to $0.8$~dB by re-marking bits via updated reliabilities over a certain number of iterations. In addition, a voting strategy based on HRBs was also proposed for PCs to determine whether a BDD output is a miscorrection or not~\cite{LiICCC2019SA}. In this method, once the number of HRBs involved in the suspected errors (detected by BDD) exceeds a threshold, the decoding result will be regarded as a miscorrection.

Most of the work on FEC decoders (including the work above) is tested by simulations with floating-point computation. However, to reduce power and cost, practical implementations, e.g., using field programmable gate arrays (FPGAs), convert the floating-point numbers to fixed-point numbers. The floating-point to fixed-point conversion, which can be seen as a quantization process, will result in a finite precision representation of the messages, which can cause a performance loss. Therefore, when hardware implementation of the FEC decoders is considered, the quantization of the channel soft information is an important issue to be considered.

Existing works on the effect of soft information quantization mainly focus on the FEC codes that use SD decoders, e.g., LDPC~\cite{Quant2009,ZhangTCOM2009,Quant_LDPC2010,FabianJLT2019}, turbo codes~\cite{Quant_Turbo2002}, and polar codes~\cite{PolarSCQuant2013,PolarAlexios2014}. Typically, 6-bit quantization is considered as the best trade-off between performance and complexity for SD-decoders.
In the context of SA-HD decoders, the authors of~\cite{Alireza2020} have briefly evaluated the effect of LLR quantization on the so-called iBDD-CR algorithm for PCs. The results in~\cite{Alireza2020} show that iBDD-CR can tolerate 3-bit LLR quantization with $0.07$~dB performance loss. In addition, the FPGA emulation with 6-bit LLR representation of the concatenated HD-based SCC and SD-based Hamming code (proposed for 400G-ZR), reveals an error flare\cite[Fig. 11]{CaiYi-JLT2018}. This is not observed in the floating-point simulations~\cite[Fig. 11]{CaiYi-JLT2018}. Fortunately, it is shown in~\cite[Fig. 12]{CaiYi-JLT2018} that increasing the decoding window size of SCCs can effectively remove the error flare in the fixed-point FPGA implementations. More details about the hardware implementation of the 400G-ZR FEC codes with quantized channel soft information have been recently given in~\cite{TruhachevTCS2020}.

In this paper, a new SA-HD decoder based on an improved SABM (iSABM) algorithm is proposed for SCCs, which we call iSABM-SCC. The main motivation is to make the iSABM-SCC decoder hardware-friendly yet still providing considerable error-correcting performance gains.
To achieve this, the idea of sorting bits by reliability (required in the SABM-SCC decoder for marking HUBs) is abandoned. This novel iSABM-SCC decoder uses two reliability thresholds to classify the bits into three types: HRBs, HUBs, and uncertain bits (UBs). To increase gains, the iSABM algorithm randomly selects HUBs for flipping and tackles more SCC blocks (rather than only the last $2$ SCC blocks used by the SABM-SCC) within a window.
The main contributions of this paper are two: {(i) a novel iSABM-SCC decoder is proposed, and its performance under different modulation formats and error-correcting capabilities of the component codes is studied;} (ii) the impact of reliability quantization on the performance of the iSABM-SCC decoder as well as the SABM-SCC decoder is shown. Numerical results show that the achieved additional gains of the iSABM-SCC decoding can be up to $0.91$~dB with respect to standard SCCs, while the performance loss caused by $1$-bit reliability quantization is $0.25$~dB.

The remainder of the paper is organized as follows. In Sec.~II, we present the system model and review the previously proposed SABM-SCC decoder. {In Sec.~III, we introduce the newly proposed iSABM-SCC decoder and show its performance with idealized three-level channel reliabilities. The impact of reliability quantization on the performance of the iSABM-SCC and SABM-SCC decoders is analyzed in Sec.~IV.} The decoding complexity is discussed in Sec.~V. Finally, Sec.~VI concludes this paper.

\section{System Model, SCCs and SABM-SCC Decoder}
\subsection{System Model}
\begin{figure}[!tb]
\includegraphics[width=0.5\textwidth]{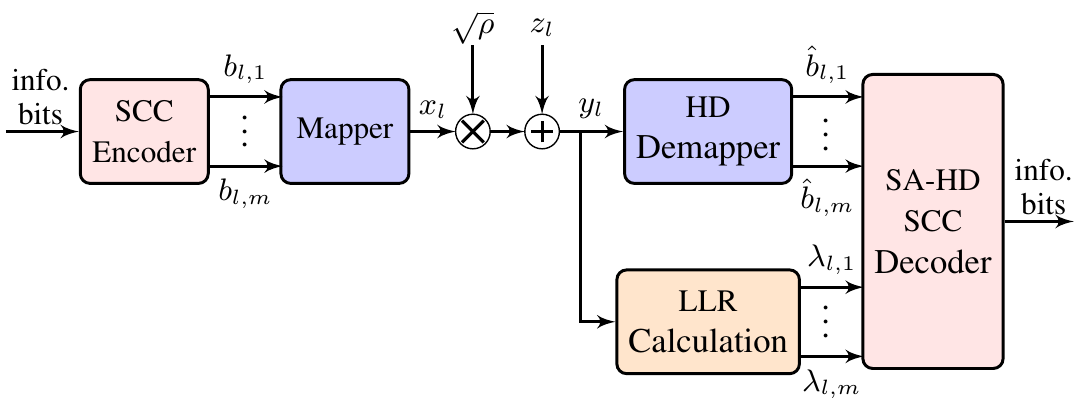}
\vspace{-2em}
\caption{System model considered in this paper.}
\label{fig:model}
\end{figure}

Fig.~\ref{fig:model} shows the system model considered in this paper.
Information bits are encoded into coded bits $b_{l,1},\ldots,b_{l,m}$ by an SCC encoder and then mapped to symbols $x_{l}$ taken from an equally-spaced $M$-ary Pulse Amplitude Modulation (PAM) constellation $\mathcal{S}=\{s_{1},s_{2},\ldots,s_{M}\}$ with $M=2^m$ points, where $l$ is the discrete time index, $l=0,1,2,\ldots$. The bit-to-symbol mapping is the binary reflected Gray code. The received signal is ${y_{l}}=\sqrt{\rho}{x_{l}}+{z_{l}}$, where ${z_{l}}$ is zero-mean unit-variance additive complex white Gaussian noise (AWGN) and $\sqrt{\rho}$ is the channel gain.

Based on the received signal $y_l$, the HD-based demapper will estimate the coded bits $\hat{b}_{l,1},\ldots,\hat{b}_{l,m}$, which are then fed to the SA-HD SCC decoder. At the same time, the receiver calculates the LLR value $\lambda_{l,k}$ for each bit, defined as ~\cite[eq.~(3.50)]{AlexBook2015}
\begin{equation}\label{LLR}
   \lambda_{l,k}=\sum_{b \in \{0,1\}} (-1)^{\bar{b}} \log\sum_{i \in \mathcal{I}_{k,b}} \textrm{exp}\left(-\frac{(y_{l}-\sqrt{\rho}s_{i})^{2}}{2}\right),
\end{equation}
with $k=1,\ldots,m$, and where $\bar{b}$ denotes bit negation. In \eqref{LLR}, the set $\mathcal{I}_{k,b}$ enumerates all the constellation points in $\mathcal{S}$ whose $k$th bit $c_{i,k}$ is $b$, i.e., $\mathcal{I}_{k,b}\triangleq \{i=1,2,\ldots,M: c_{i,k}=b\}$. The LLRs are then provided to the SA-HD SCC decoder.

Within the SA-HD SCC decoder, HD decoding is performed to decode the HD-estimated bits with the help of channel LLRs. The SA-HD SCC decoding can be performed in multiple ways, such as SABM-SCC~\cite{YiISTC2018,YiTCOM2019}, BEE-SCC~\cite{AlirezaSCC2020}, etc. In what follows, we will review SCCs and the SABM-SCC decoder.

\subsection{Staircase Codes}

An SCC consists in a concatenation of binary matrices $\boldsymbol{B}_{i}\in\{0,1\}^{w \times w}$, $i=0, 1, 2,\ldots$, which can be graphically represented by a staircase. For $\forall i>1$, each row of the matrix $[\boldsymbol{B}^{T}_{i-1} \boldsymbol{B}_{i}]$ is a valid codeword in a component code $\mathcal{C}$, where $\boldsymbol{B}^{T}_{i-1}$ is the matrix transpose of $\boldsymbol{B}_{i-1}$. In this paper, we consider BCH codes with parameters of $(n_{c}, k_{c}, t)$ as the component codes $\mathcal{C}$, where $n_{c}$ is the component codeword length, $k_{c}$ is the information length, and $t$ is the error-correcting capability. The code rate of SCC is defined as $R = 2k_{c}/n_{c}-1$, while the size of $\boldsymbol{B}_{i}$ is $w=n_{c}/2$.

Standard SCCs are decoded by using a sliding window covering $L$ received SCC blocks $\{\boldsymbol{Y}_{i}, \boldsymbol{Y}_{i+1},\ldots, \boldsymbol{Y}_{i+L-1}\}$ (as the red area shown in Fig.~\ref{fig:SABM}), where $\boldsymbol{Y}_{i}$ corresponds to the transmitted SCC block $\boldsymbol{B}_{i}$. Within the window, BDD is used to iteratively decode each received component word from the bottom left to the top right. BDD is very simple, however, it can only handle the cases with $t$ or less than $t$ errors. In the case of more than $t$ errors in the received
component word, either a decoding failure or a miscorrection occurs. Miscorrection is a situation where BDD finds a codeword in the codebook with up to $t$ different bits from the received one, but this codeword does not correspond to the transmitted one. That is, the received component word is erroneously decoded to another codeword in $\mathcal{C}$. Miscorretion is known to degrade the performance, especially in the iterative decoding process.

\subsection{The SABM-SCC Decoder}

\begin{figure*}[!tb]
\centering
\includegraphics[width=0.88\textwidth]{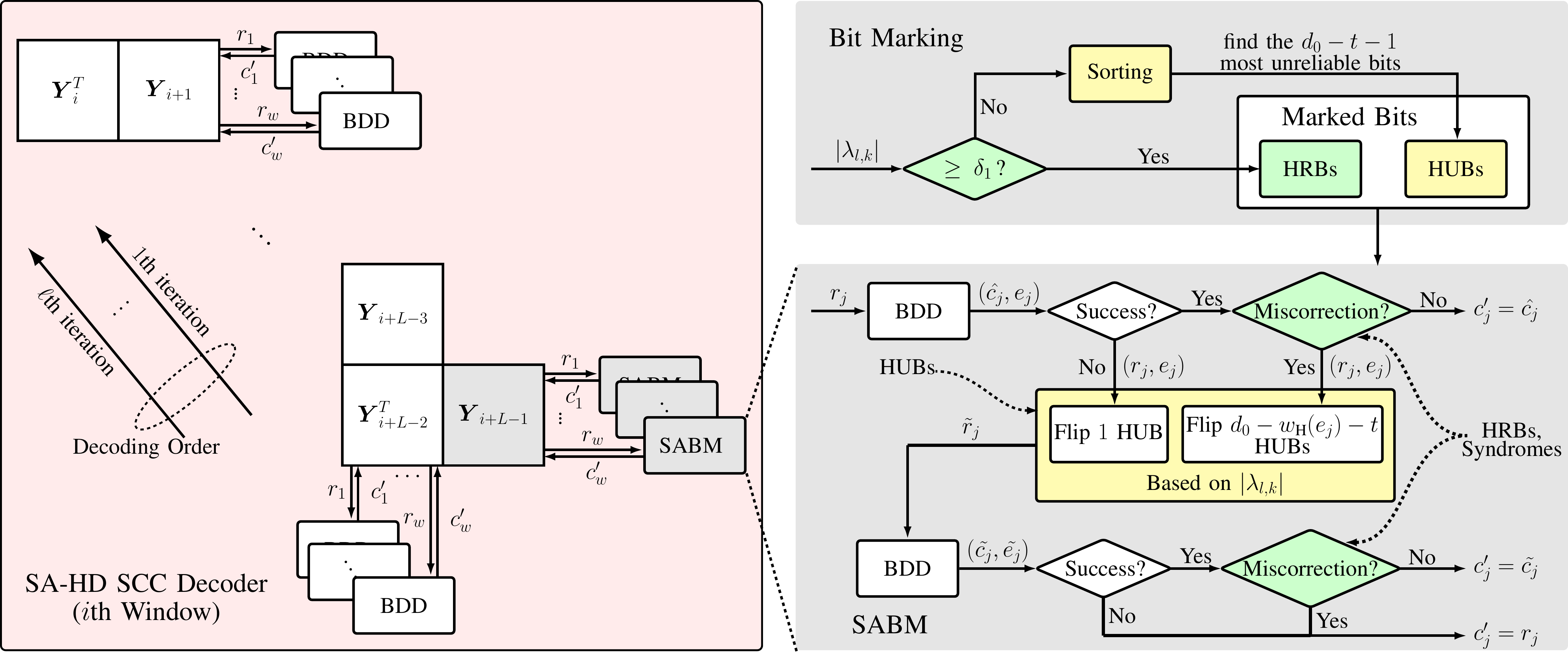}
\caption{Flow chart (left) of the SABM-SCC decoder in the $i$th window, where $L$ is the window size. $r_j$ is a received component word taken from one of the rows or columns of two neighbor SCC blocks, and $c'_j$ is the corresponding output of the component decoder, $j=1, \ldots, w$. The right figures show the workflows of bit marking and SABM decoding.}
\label{fig:SABM}
\end{figure*}

To improve the performance of SCCs, the SABM-SCC decoder is proposed in~\cite{YiISTC2018,YiTCOM2019}. Fig.~\ref{fig:SABM} shows the flow chart of the SABM-SCC decoding in the $i$th window. For two neighboring SCC blocks, $w$ component decoders are typically performed in parallel to decode the $w$ component words $r_j$, $j=1, \ldots, w$, i.e., the $w$ rows or columns of the two neighbor SCC blocks. We treat the $w$ component decoders as a group. Instead of $L-1$ groups of BDDs in the standard SCC decoder, the SABM-SCC decoder uses $L-2$ groups of BDDs and one group of ``special'' BDDs, i.e., SABMs, at each iteration.

The SABM algorithm is based on the idea of marking bits. It uses the absolute LLR value $|\lambda_{l,k}|$ to represent the reliability of a bit $\hat{b}_{l,k}$ (a higher value of $|\lambda_{l,k}|$ indicates a more reliable bit). Based on this, the SABM-SCC decoder marks the HD-estimated bits as HRBs, HUBs, or UBs. As shown in the top right of Fig.~\ref{fig:SABM}, a threshold $\delta_1$ is used to classify HRBs, i.e., to fall within the HRB class, $|\lambda_{l,k}|$ should be larger than $\delta_1$. To mark HUBs, the SABM-SCC decoder needs to sort the reliabilities with the indices corresponding to the bits in each row of an SCC block. The aim of this is to find the sorted $d_0-t-1$ bits (out of the $w$ bits) with the smallest $|\lambda_{l,k}|$ values, where $d_0$ is the minimum Hamming distance of the component code $\mathcal{C}$. The sorted $d_0-t-1$ bits in each row are the HUBs, while the UBs are the bits that are neither HRBs nor HUBs. With the marked information in $\boldsymbol{Y}_{i+L-1}$ (where most errors are located), SABM is performed to decode the received component words in the last $2$ SCC blocks, i.e., $[\boldsymbol{Y}^{T}_{i+L-2} \boldsymbol{Y}_{i+L-1}]$, within a window. 

The bottom right of Fig.~\ref{fig:SABM} shows the workflow of the SABM algorithm to decode a received component word $r_j$.
Unlike standard SCC decoding, the SABM decoder does not trust the output of BDD unconditionally. It will detect whether or not the output of BDD is a miscorrection, if BDD declares success. One criterion is that no HRBs in $\boldsymbol{Y}_{i+L-1}$ should be flipped. As the successfully decoded component codeword has a zero syndrome, the errors detected by BDD in $\boldsymbol{Y}_{i+L-2}$ should not be in conflict with zero-syndrome component codewords in $[\boldsymbol{Y}^{T}_{i+L-3} \boldsymbol{Y}_{i+L-2}]$ either. Only when the two criteria are satisfied, the output of BDD will be accepted. Otherwise, it will be regarded as a miscorrection and be rejected.

For the miscorrections, the SABM decoder will flip the most unreliable $d_{0}-w_{\text{H}}(e_j)-t$ bits in $r_j$ in $\boldsymbol{Y}_{i+L-1}$, where $e_j$ is the error pattern detected by BDD and $w_{\text{H}}(\cdot)$ is the Hamming weight. For the BDD failures, the SABM decoder will flip the most unreliable bit in $r_j$ in $\boldsymbol{Y}_{i+L-1}$. The intuition here is that bits with the lowest reliabilities are the most likely channel errors. In some cases, bit-flipping (BF) will make the resulted sequence $\tilde{r}_j$ close enough to the transmitted codeword $c_j$, i.e.,  $d_\text{H}(\tilde{r}_j,c_j)=t$, where $d_{\text{H}}(\cdot,\cdot)$ represents the Hamming distance. Thus, when the second BDD attempt is performed, the residual errors in $\tilde{r}_j$ can be corrected. In case that BF results in a wrong decision, i.e., the flipped bits are not errors, miscorrection detection will be performed to make a final check if the decoding succeeds. The achieved performance and increased complexity of the SABM-SCC decoder were discussed in~\cite{Alex_OECC2019,YiTCOM2019}.

\section{The iSABM-SCC Decoder and Its Performance}

Although the SABM-SCC decoder was shown to achieve considerable gains, it still presents some shortfalls which leave margins for improvement. In this section, we propose the iSABM-SCC decoder, which is shown in Fig.~\ref{fig:iSABM}. The details of the iSABM-SCC decoder are given in what follows.

\subsection{The Proposed iSABM-SCC Decoder}
As shown in the top right of Fig.~\ref{fig:SABM}, sorting is required for marking HUBs. On the one hand, as pointed out in~\cite[Sec. 2.2]{Alex_OECC2019}, the sorting process plays a significant role in the complexity increase of the SABM-SCC decoding, as every row of an SCC block needs to be sorted. To reduce the complexity overhead, sorting bits should be avoided if possible. On the other hand, for hardware implementation, when the reliability is represented using a finite number of bits, reliabilities with close floating-point values will be quantized to the same fixed-point value. This will make it hard to find a unique set of sorted $d_0-t-1$ HUBs out of the $w$ bits in each row of an SCC block, in particular when the reliabilities are coarsely quantized. As a result, the sorting process can be heavily affected by a potential ordering ambiguity, which also depends on the exact hardware sorting network that is used. {We will explain more about this later with simulation results in Sec. IV-B.}

\begin{figure*}[!tb]
\centering
\includegraphics[width=0.94\textwidth]{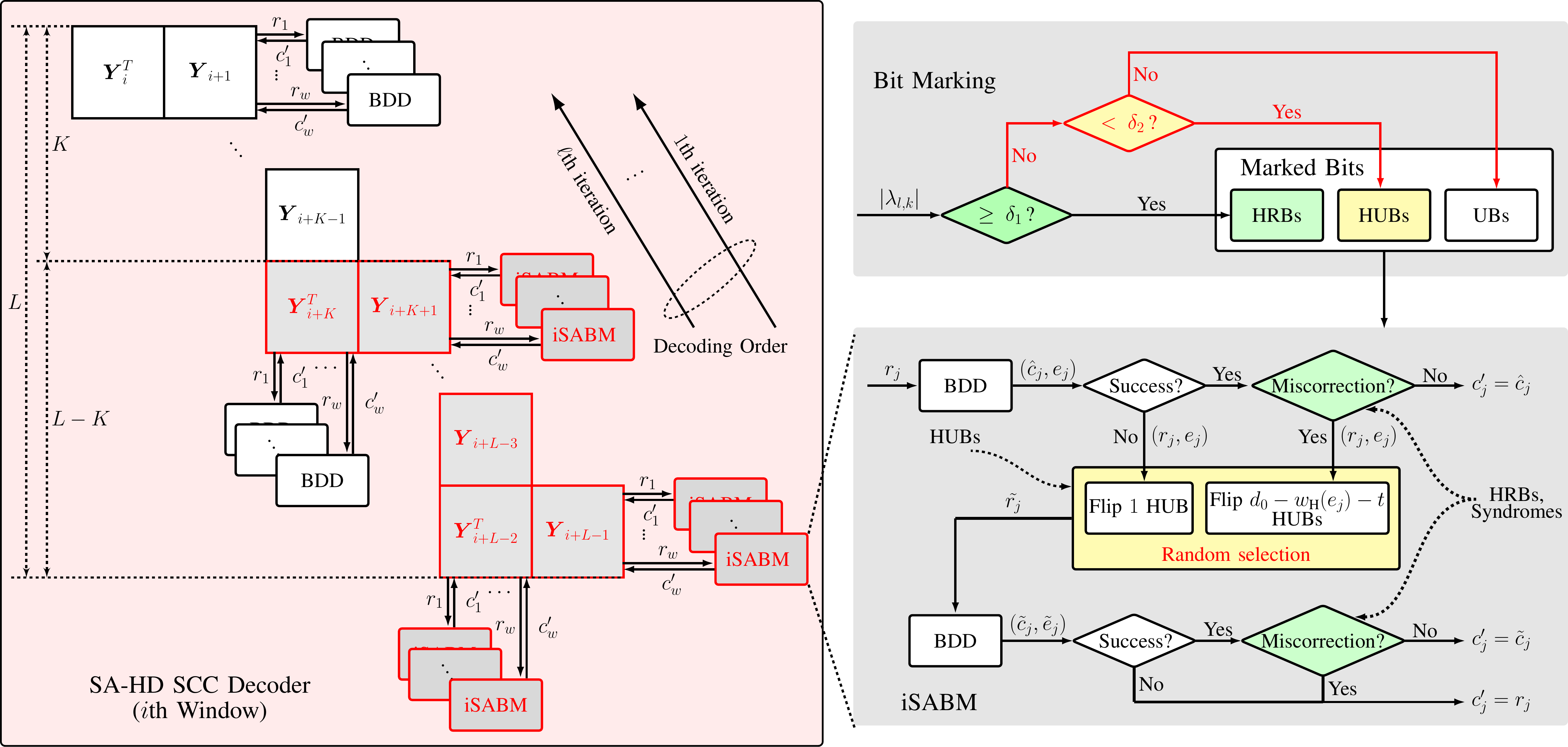}
\caption{Flow chart (left) of the proposed iSABM-SCC decoder in the $i$th window. The right figures show the workflows of bit marking and iSABM decoding. The red highlighted parts show the differences between the iSABM-SCC and SABM-SCC decoders (see Fig.\ref{fig:SABM}).}
\label{fig:iSABM}
\end{figure*}

To make it simple and suitable for coarse reliability quantization, we focus on improving the bit marking strategy for the iSABM-SCC decoder. Similarly to the process of marking HRBs, a second reliability threshold $\delta_2$ is introduced to classify HUBs instead of sorting bit reliabilities. This is shown with a red decision block at the top right of Fig.~\ref{fig:iSABM}. According to the value of $|\lambda_{l,k}|$, the marking result for a bit $\hat{b}_{l,k}$ is given by

\begin{equation}\label{BitMarking}
     \begin{aligned}
     \left\{
     \begin{array}{lcl}
     \text{HRB},  &      & \text{if~} |\lambda_{l,k}| \geq \delta_1 \\
     \text{UB}, &      & \text{if~} \delta_2 \leq |\lambda_{l,k}| < \delta_1 \\
     \text{HUB},  &      & \text{if~} |\lambda_{l,k}| < \delta_2\\
     \end{array}
     \right.
     \end{aligned}.
\end{equation}

Another weakness of the SABM-SCC decoder is that the un-updated reliabilities will make some received component words have little chance to be corrected. For the decoding failures and miscorrections in the second BDD attempt, the SABM algorithm will return the received component word $r_j$. If no errors in $r_j$ are corrected by the following component decodings, the first BDD attempt for $r_j$ in the next iteration will also be a failure or a miscorrection (as the number of errors in $r_j$ is still beyond the error-correcting capability of BDD). In the SABM algorithm, the most unreliable bits will then be flipped. However, as the reliabilities are not updated as iterations go on, the flipped bits are always the same as that in the previous iteration. This will make the second BDD attempt for $r_j$ suffer from the same experience with that in the previous iteration, i.e., decoding failure or miscorrection occurs.

To give more chances to decode the BDD failures and miscorrections, the iSABM-SCC decoder randomly selects HUBs for flipping.
This will result in different bit flipping trials in different iterations. The number of flipped bits is identical to that in the SABM algorithm, i.e., $d_0-t-w_\text{{H}}{(e_j)}$ and $1$ for the miscorrections and failures, respectively. In some cases, the number of HUBs may be less than that of the required bit flips. For these cases, even the decoder flips all the HUBs, the resulted sequence will not be close enough to the transmitted codewords. Therefore, the iSABM algorithm will give up BF and keep the component word unchanged. For the miscorrection detection, the iSABM algorithm follows the same rules as SABM: (i) no HRBs are flipped, and (ii) no suspected errors (detected by BDD) are in conflict with the zero-syndrome component codewords.

In addition, soft-aided decoding in the SABM-SCC decoder only tackles the last $2$ SCC blocks, i.e., $[\boldsymbol{Y}^{T}_{i+L-2} \boldsymbol{Y}_{i+L-1}]$, to keep the complexity increase as low as possible. Moreover, the soft information in the last SCC block $\boldsymbol{Y}_{i+L-1}$ is used in the SABM algorithm, while the bits in the other SCC block $\boldsymbol{Y}_{i+L-2}$ get no benefit from the soft information. Although this limits the complexity increase, the performance improvement is limited as well.

To obtain more gains, more soft information in the SCC blocks can be utilized. As shown in Fig.~\ref{fig:iSABM} (left), the iSABM-SCC decoder uses the soft information in the last $L-K$ SCC blocks, $K=0,\ldots,L-1$. For $K < L-1$, the iSABM-SCC decoder performs $K$ groups of BDDs and $L-K-1$ groups of iSABMs at each iteration. In particular, $K=0$ means all the SCC blocks are tackled by the iSABM algorithm. For $K=L-1$, it is a special case that iSABM performs with half of marked bits in the component codewords, as only the soft information in $\boldsymbol{Y}_{i+L-1}$ is used. In this case, the iSABM-SCC decoder performs $L-2$ groups of BDDs and one group of iSABMs at each iteration. This is similar to the SABM-SCC decoder. The only difference is that iSABM randomly selects HUBs for flipping, while SABM always selects the HUBs with least reliabilities for flipping.

As the iterations go on, the marked information is not updated. This is mainly because the bits in the correctly decoded component codewords will be protected by the zero syndromes. Once a BDD output is in conflict with zero-syndrome component codewords, it will be regarded as a miscorrection and be rejected by the iSABM algorithm. Therefore, the correctly decoded bits have little chance to be wrongly flipped in the next decodings.

\subsection{Numerical Results}

In this section, the performance of the iSABM-SCC decoder is evaluated using numerical simulations over an AWGN channel. In the simulations, we first consider SCCs with BCH $(256,239,2)$ component code. It is extended by adding an additional parity bit at the end of standard BCH code of $(255,239,2)$. The resulting SCC code rate is $R=0.87$. The number of iterations is $\ell=7$, while the decoding window size is $L=9$. The HRB and HUB marking thresholds are $\delta_1=10$ and $\delta_2=2.5$, respectively. The two thresholds are numerically optimized by testing different values at a signal-to-noise ratio (SNR) of $6.45$ dB for 2-PAM to have the best BER performance.

\begin{figure}[!tb]
\includegraphics[width=0.48\textwidth]{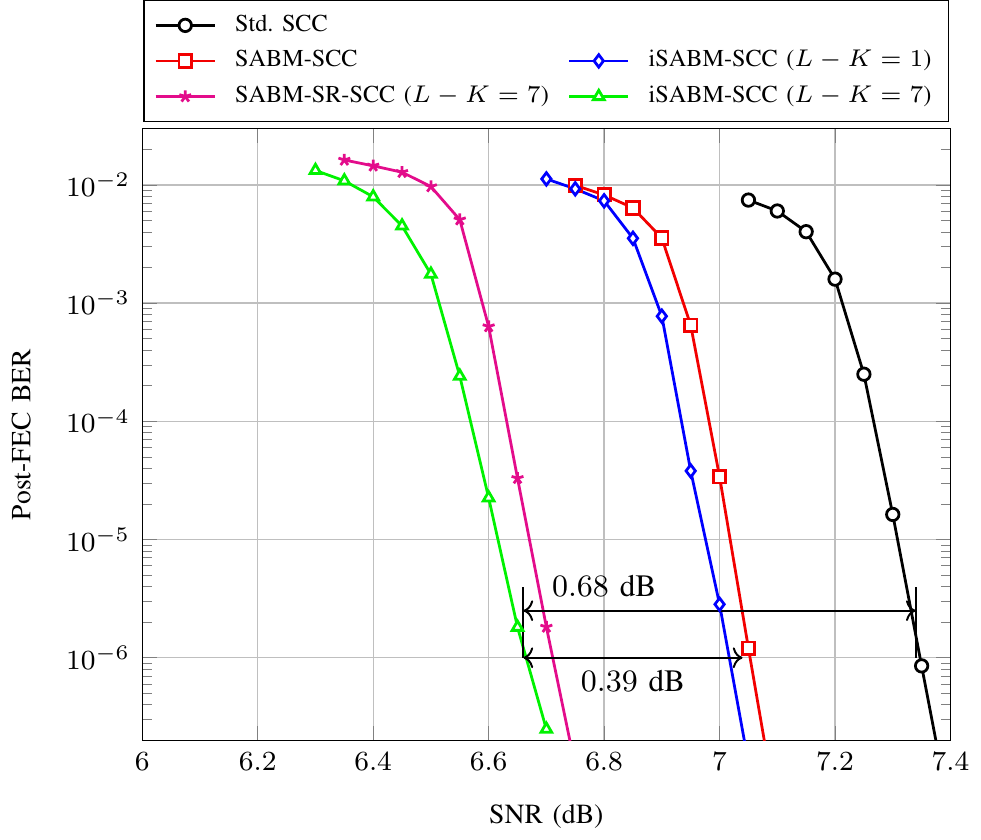}
\caption{Post-FEC BER vs. SNR for SCCs with BCH $(256,239,2)$ component code. The resulting SCC code rate is $R=0.87$ and the modulation format is 2-PAM.}
\label{fig:iSABM_t=2-2PAM}
\end{figure}

Fig.~\ref{fig:iSABM_t=2-2PAM} shows the BER performance of the iSABM-SCC decoding for 2-PAM against SNR. For comparison, it also includes two performance baselines: SABM-SCC decoding (red curve) and standard SCC decoding (black curve). The magenta curve is the performance of the SABM-SR algorithm for SCCs, which is extended from the work in~\cite{Gabriele_ECOC2019}. We will refer it as SABM-SR-SCC.

First of all, Fig.~\ref{fig:iSABM_t=2-2PAM} shows that iSABM-SCC with {$L-K=1$ (i.e., $K=8$)} outperforms SABM-SCC, when both of them use the soft information in the last SCC block of a window. This benefit comes from the random selection of HUBs for flipping. As explained in the third and fourth paragraphs of Sec. III-A, random selection of HUBs can result in different bit flipping trials in different iterations. This gives more chances to the iSABM-SCC decoder to find the channel errors for flipping, when the decoding in the previous iteration is a miscorrection or a failure. However, the small gap between the red and blue curves indicates that random BF is not enough for iSABM-SCC to obtain large gains.

To have more gains, soft-aided decoding combined with random BF is performed over more SCC blocks. The green curve in Fig.~\ref{fig:iSABM_t=2-2PAM} shows the BER performance of iSABM-SCC with {$L-K=7$ (i.e., $K=2$)}. In this case, iSABM-SCC uses soft information in the last $7$ SCC blocks of a window. As can be seen, iSABM-SCC can outperform SABM-SCC by up to $0.39$~dB, while the overall additional gain is up to $0.68$~dB when compared to standard SCC at a post-FEC BER of $10^{-6}$. It is found that this is the optimal performance of the iSABM-SCC decoder. Tackling beyond the $7$ blocks (out of the $9$ SCC blocks of the window) will degrade the decoding performance. The intuition for this is due to the inaccurate marked information in the first two blocks after multiple decodings in the previous windows.

When compared to the SABM-SR-SCC decoder, the iSABM-SCC decoder with {$L-K=7$} shows a slightly better performance. The optimization of the scaling weights for LLR update follows the same method in~\cite{Gabriele_ECOC2019}. Since each bit is decoded twice by the component decoder at each iteration (one row component decoding and one column component decoding (see Fig.~\ref{fig:iSABM}(left)), the scaling weight vector $\boldsymbol{v}$ for LLR update contains $14$ elements in the case of $\ell=7$ iterations. The optimized scaling weight vector for the $14$ elements  we used is $\boldsymbol{v}=[8.6, 8.7, 8.4, 9, 9.7, 9.9, 10.5, 11.4, 12, 11.2, 11.5, 12, 13.4, 13.5]$. The $i$th element in $\boldsymbol{v}$ corresponds the optimum scaling weight for the $i$th decoding of the bits.

{The achieved performance for SABM-SR-SCC here may be suboptimal, as the optimization of the scaling weights did not consider the different reliability distribution among the SCC blocks of a window. Differently from PCs, where the component decoding is performed iteratively within the same block, the decoding of SCCs performs using a sliding window. The blocks in the front (i.e., in the top right) of the window are decoded more times. As a result, the closer to the front of the SCC block is, the more reliable of the decoding will be. Therefore, when the SABM-SR algorithm is extended from PCs to SCCs, the scaling weights may be different for each SCC block needs to be considered to have optimum performance~\cite{AlirezaSCC}. However, this will increase the dimensions of the scaling weight vector, and thus make the optimization process very time-consuming if Monte-Carlo simulations are used. }

\begin{figure}[!tb]
\includegraphics[width=0.48\textwidth]{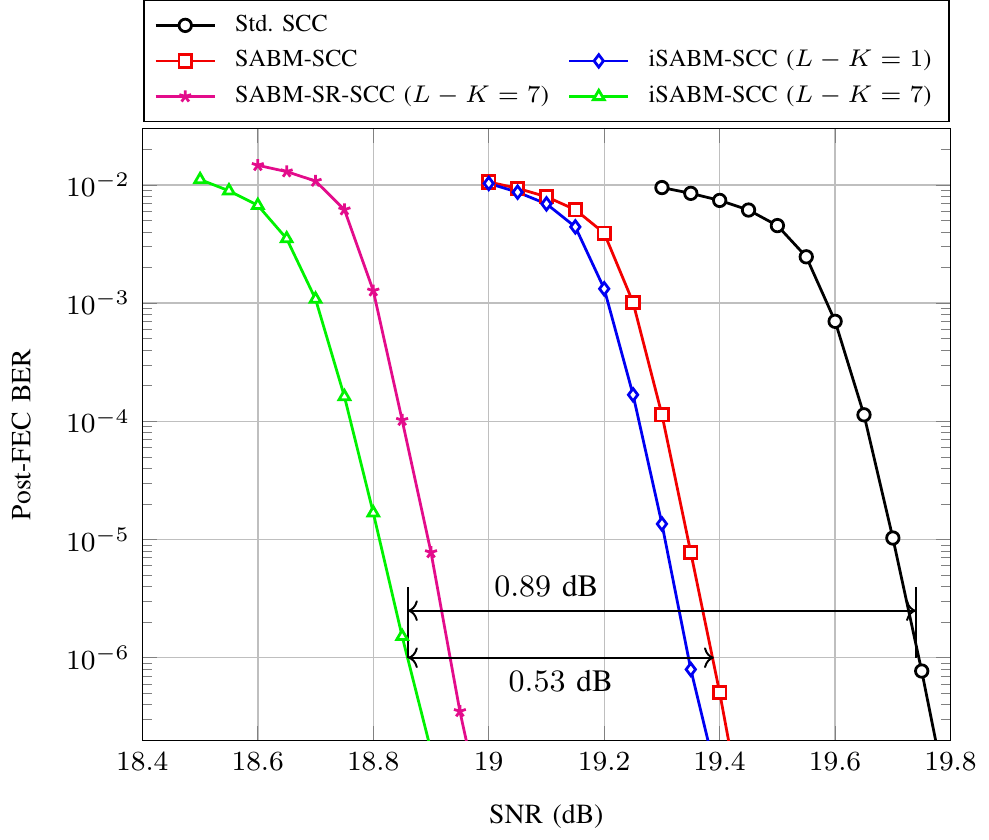}
\caption{Post-FEC BER vs. SNR for SCCs with BCH $(256,239,2)$ component code. The resulting SCC code rate is $R=0.87$, and  the modulation format is 8-PAM.}
\label{fig:iSABM_t=2-8PAM}
\end{figure}

Fig.~\ref{fig:iSABM_t=2-8PAM} shows the BER performance of iSABM-SCC decoding for 8-PAM. It is observed that the gains are somewhat higher than that for $2$-PAM. The achieved additional gain of iSABM-SCC can be up to $0.53$~dB with respect to SABM-SCC, while the overall improvements are up to $0.89$~dB when compared to standard SCC.

\begin{figure}[!tb]
\includegraphics[width=0.5\textwidth]{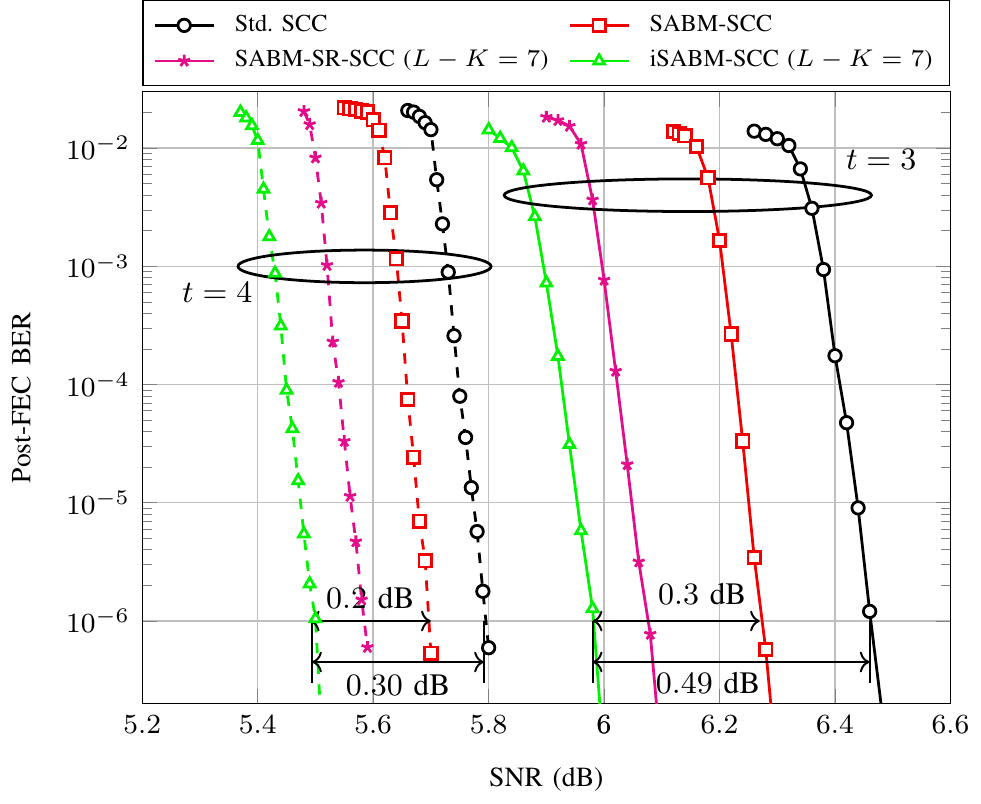}
\caption{Post-FEC BER vs. SNR for SCCs with BCH $(256,231,3)$ and $(256,223,4)$ component codes. The resulting SCC code rates are $R=0.80$ and $R=0.74$, respectively, and the modulation format is 2-PAM. }
\label{fig:iSABM_t_3_4}
\end{figure}

As component codes with $t=3$ and $t=4$ are more interesting in practice ($t=3$ has been recommended in the ITU standards for OTNs~\cite{G709.2,G709.3,OIF400G}), we further investigate the performance of iSABM-SCC with BCH $(256,231,3)$ and BCH $(256,223,4)$ component codes.
They are 1-bit extended codes based on standard BCH codes with parameters of $(255,231,3)$ and $(255,223,4)$, respectively.
The resulting SCC code rates are $R=0.80$ and $R=0.74$, respectively. Fig.~\ref{fig:iSABM_t_3_4} shows the simulation results for 2-PAM. To avoid too many curves, only the results of iSABM-SCCs with {$L-K=7$} are shown. It can be seen that even with larger $t$, iSABM-SCC still outperforms SABM-SCC and SABM-SR-SCC. Specifically, iSABM-SCC offers additional gains of $0.30$~dB and $0.20$~dB for $t=3$ and $t=4$ with respect to SABM-SCC, respectively. The overall additional gains are increased to $0.49$~dB and $0.30$~dB, respectively, when compared to standard SCCs.

To compare with the latest BEE-SCC decoder proposed in~\cite{AlirezaSCC2020}, SCCs with BCH $(254,230,3)$ component code are also considered for $2$-PAM and $16$-PAM. The BCH code is obtained by shortening $1$ information bit based on BCH $(255,231,3)$. For the sake of fairness, we use the same parameters as~\cite[Fig. 10]{AlirezaSCC2020} and ~\cite[Fig. 12]{AlirezaSCC2020}, i.e., $L=7$ window size and $\ell=10$ iterations. For iSABM-SCC, we use {$L-K=5$, i.e., $K=2$}. A random interleaver is employed within each SCC block. The results in Fig.~\ref{fig:BCH(254,230,3)} show that iSABM-SCC presents a slightly better BER performance than BEE-SCC. With respect to standard SCCs, the additional gains for $2$-PAM and $16$-PAM are $0.58$~dB and $0.91$~dB, respectively. A complexity comparison between iSABM-SCC and BEE-SCC will be presented in Sec. V.

\begin{figure}[!tb]
\includegraphics[width=0.5\textwidth]{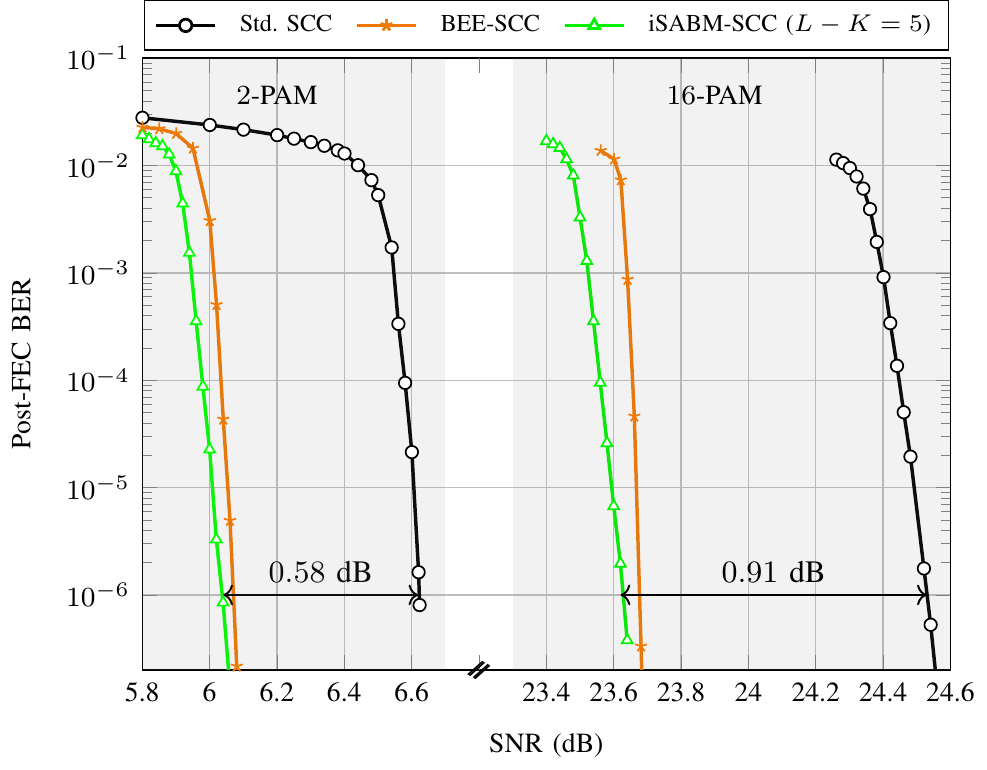}
\caption{Post-FEC  BER  vs.  SNR  for  SCCs  with BCH $(254,230,3)$ component  code for $2$-PAM and $16$-PAM. The resulting SCC code rate is $R=0.811$.
}
\label{fig:BCH(254,230,3)}
\end{figure}

\section{Effect of Reliability Quantization on the {SABM- and iSABM-SCC} Decoders}

In hardware implementations of 400G-ZR receivers, the channel output $y_l$ commonly uses $6$-bit or $7$-bit representation~\cite{YiTCOM2019,TruhachevTCS2020}. As the LLR calculation operates with the quantized value of $y_l$, the LLRs are naturally quantized. Since $6$-bit or $7$-bit representation is almost as good as a floating-point representation, its effect on the performance can be ignored. Therefore, this paper assumes that the LLRs are calculated using floating-point arithmetic. Under this assumption, this section will study the effect of fixed-point representation of the floating-point values of $|\lambda_{l,k}|$, i.e., reliability quantization, on the performance of SABM- and iSABM-SCC decoders.

\subsection{Reliability Quantization}

\begin{figure}[!tb]
\centering
    \label{subfig:SABMQuantization}
\includegraphics[width=0.5\textwidth]{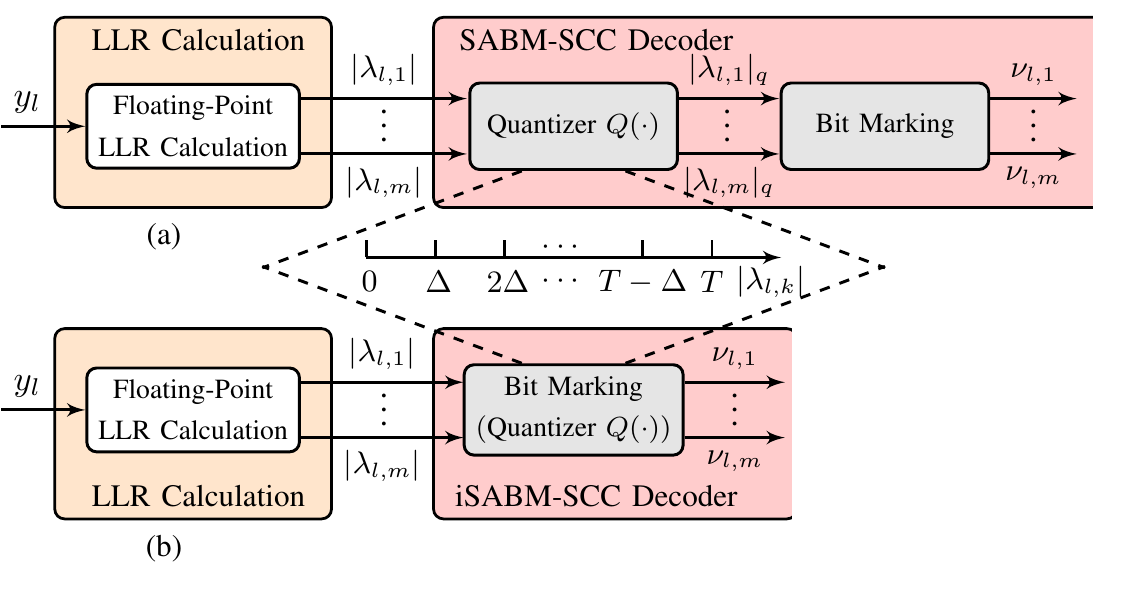}
  \hspace{0ex}
\caption{Reliability quantization we considered in the (a) SABM-SCC and (b) iSABM-SCC decoders. $T$ is the saturation threshold of the quantizer $Q(\cdot)$ with resolution of $\Delta$.}
\label{fig:LLR_quantization}
\end{figure}

Fig.~\ref{fig:LLR_quantization} shows the reliability quantization we considered in the SABM- and iSABM-SCC decoders. The floating-point value of $|\lambda_{l,k}|$ is calculated from the floating-point channel output $y_l$ according to \eqref{LLR}, where $|\lambda_{l,k}| \in [0,+\infty)$. To perform sorting, the SABM-SCC decoder needs to store the reliabilities $|\lambda_{l,k}|$. In hardware implementation, this will make the SABM-SCC decoder encounter the problem of fixed-point representation of $|\lambda_{l,k}|$. Here, a $q$-bit quantizer $Q(\cdot)$ is used to convert $|\lambda_{l,k}|$ into a fixed-point value $|\lambda_{l,k}|_q$. With $|\lambda_{l,k}|_q$, bit marking unit in the SABM-SCC decoder sorts the bits $\hat{b}_{l,k}$, and classifies them into HRBs, UBs, and HUBs (see the top right of Fig.~\ref{fig:SABM}). Contrary to SABM-SCC, iSABM-SCC does not require storing channel relibilities as no sorting is needed. Instead, the values $|\lambda_{l,k}|$ are directly sent to the bit marking unit for classifying the bits into HRBs, HUBs, and UBs by using the two reliability thresholds $\delta_1$ and $\delta_2$ (see the top right of Fig.~\ref{fig:iSABM}). In this sense, bit marking in iSABM-SCC is equivalent to a reliability quantization process.

Bit marking in the iSABM-SCC decoder shown in Sec. III can be implemented using a three-level nonuniform quantizer, which has two decision boundaries: $\delta_1=10$ and $\delta_2=2.5$. However, in hardware, \emph{uniform} quantization is simply and readily implemented. Therefore, we further consider a $2$-bit uniform quantization for iSABM-SCC to classify the bits $\hat{b}_{l,k}$ into HRBs, HUBs, and UBs. As $|\lambda_{l,k}|$ is nonnegative, the quantizer is unsigned.
In terms of the  memory and power consumption, $1$-bit quantization is extremely interesting for practical applications. Therefore, we will also study the performance of iSABM-SCC as well as SABM-SCC with 1-bit reliability quantization.

The quantization scheme is shown in the middle of Fig.~\ref{fig:LLR_quantization}, where $T$ is the saturation threshold of the $q$-bit quantizer with resolution of $\Delta$, with $\Delta=T/2^q$. The corresponding law is given by
\begin{equation}\label{Eq:QuantEquation}
     \begin{aligned}
     |\lambda_{l,k}|_q &\triangleq Q(|\lambda_{l,k}|)=\left\{
     \begin{array}{ll}
    \lfloor \frac{ |\lambda_{l,k}|}{\Delta}\rfloor\Delta,         & 0 \leq  |\lambda_{l,k}| < T-\Delta  \\
     T-\Delta,       &  |\lambda_{l,k}| \geq T-\Delta \\
     \end{array}
     \right.
     \end{aligned},
\end{equation}
where $\lfloor \cdot \rfloor$ is the floor function that returns the largest integer number smaller than or equal to its argument.

In order to minimize the performance loss, quantization should not change the reliability levels of the bits with respect to that without quantization. To meet this requirement as much as possible, one possible way is to make the HRB threshold $\delta_1$ be one of the decision boundaries of the quantizer\footnote{This will ensure that the HRBs are immune from the reliability quantization, as their quantized reliabilities are still larger than or equal to the HRB threshold $\delta_1.$}. Another reason for this choice is that the iSABM algorithm as well as the SABM algorithm is found to be very sensitive to the change of HRBs.

According to~(\ref{Eq:QuantEquation}), we know that $m\Delta$, $m=1,\ldots, 2^{q}-1$, are the decision boundaries of the quantizer. Therefore, $\delta_1=m\Delta=mT/2^q$ is considered. Further, $m$ takes the maximum value, i.e., $m=2^{q}-1$, or in other words, $\delta_1=T-\Delta$. This will result in the smallest value of $\Delta$ for a given $\delta_1$, thus minimizing quantization error.
Therefore, we finally have
\begin{equation}
 T=\delta_1 \frac{2^q}{2^q-1}.
\label{Eq:QuantizationBound}
\end{equation}
According to (\ref{Eq:QuantizationBound}), if $\delta_1=10$, 2-bit quantization results in $T=40/3$ and $\Delta=10/3$, while 1-bit quantization results in $T=20$ and $\Delta=10$.

\begin{figure}[!tb]
\centering
 \subfigure[]{
    \label{subfig:woquantizationPDF}
\includegraphics[width=0.48\textwidth]{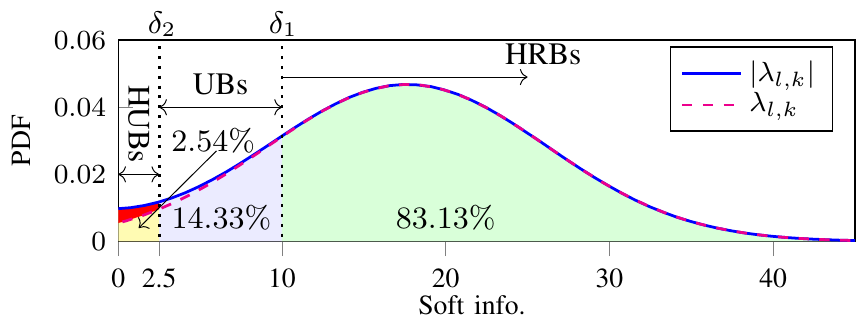}}
  \vspace{-1ex}
   \subfigure[]{
    \label{subfig:2bitquantizationPDF}
\includegraphics[width=0.48\textwidth]{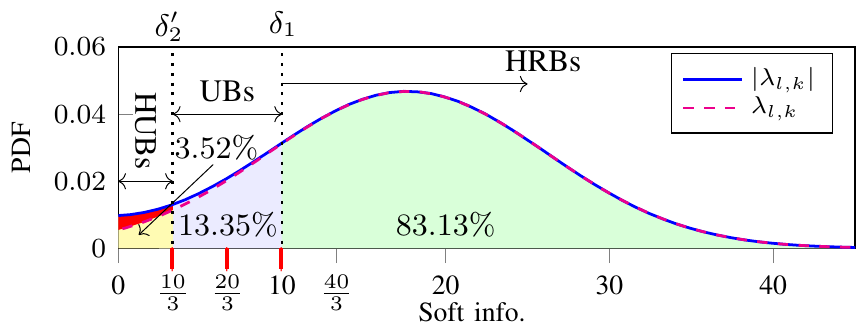}}
  \vspace{-1ex}
   \subfigure[]{
    \label{subfig:1bitquantizationPDF}
     \includegraphics[width=0.48\textwidth]{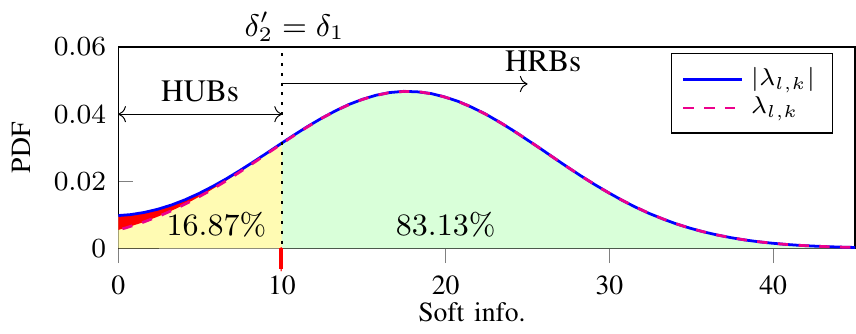}}
  \hspace{0ex}
  \vspace{-1ex}
\caption{PDFs of the reliabilities $|\lambda_{l,k}|$ and channel LLRs $\lambda_{l,k}$ (positive part) at an SNR of $6.57$~dB for 2-PAM: (a) without reliablity quantization, (b) with 2-bit reliability quantization, and (c) with 1-bit reliability quantization. }
  \vspace{-1.5ex}
\label{fig:ReliabilityPDF-1bit-2bitQuant}
\end{figure}

Fig.~\ref{fig:ReliabilityPDF-1bit-2bitQuant} shows the probability density functions (PDFs) of the reliabilities $|\lambda_{l,k}|$ and channel LLRs $\lambda_{l,k}$ (positive part) at an SNR of $6.57$~dB, which corresponds to a post-FEC BER of $10^{-3}$ for the iSABM-SCC shown in Fig.~\ref{fig:iSABM_t=2-2PAM}. The green area indicates the proportion of HRBs in the total bits both for the iSABM- and SABM-SCC decoders. The area between the blue solid line and the x-axis from 0 to $\delta_2$, i.e., red plus yellow area, indicates the proportion of HUBs in the total bits {for the iSABM-SCC decoder}. The proportion of errors in the HUBs is the ratio of the red area to the red plus yellow area, while the yellow area indicates the wrongly marked HUBs (which are correct bits).
Fig.~\ref{subfig:2bitquantizationPDF} and~\ref{subfig:1bitquantizationPDF} are the cases with $2$-bit and $1$-bit reliability quantizations. The red ticks are the decision boundaries of the quantizers.

Fig.~\ref{fig:ReliabilityPDF-1bit-2bitQuant} shows that the HRBs are not affected by the $1$-bit or $2$-bit reliability quantization we proposed. This is not the case for HUBs in the iSABM-SCC decoder. For the 2-bit quantization, the value of $|\lambda_{l,k}|$ between $0$ and $10/3$ are not recognizable anymore, as they are all assigned a quantized value of $0$. This will make the bits with $|\lambda_{l,k}|$ between $2.5$ and $10/3$ added to the HUB class in the iSABM-SCC decoder. In other words, the HUB threshold is equivalently changed to $10/3$ ($\delta'_2$ is used to indicate the new HUB threshold in Fig.~\ref{subfig:2bitquantizationPDF}). In this case, the change of the proportion of HUBs in the bits is very small, i.e., only from $2.54\%$ to $3.52\%$.
Therefore, we can expect a negligible performance loss for iSABM-SCC with this 2-bit reliability quantization. In the case of 1-bit reliability quantization, $\delta'_2=\delta_1=10$. It is equivalent to a binary decision on the bits: HUBs with $|\lambda_{l,k}| < \delta_1$ and HRBs with $|\lambda_{l,k}| \geq \delta_1$. As shown in Fig.~\ref{subfig:1bitquantizationPDF}, the number of HUBs is significantly increased from $2.54\%$ to $16.87\%$. More importantly, the proportion of errors in the HUBs becomes less, due to the greatly increased yellow area. This will lead to that BF has a higher probability to flip the wrong bits, which potentially gives a larger performances loss.

With respect to HUBs in the SABM-SCC decoder, they are always the $d_0-t-1$ bits with the smallest values of $|\lambda_{l,k}|$ in each row of a SCC block. In the example of SCC with $w=128$ and BCH $(256,239,2)$, $d_0-t-1=3$. However, Fig.~\ref{subfig:2bitquantizationPDF} and~\ref{subfig:1bitquantizationPDF} show that in the cases of 2-bit and 1-bit reliability quantizations for 2-PAM, $3.52\%$ and $16.87\%$ of bits (with the smallest $|\lambda_{l,k}|$ values), i.e., $5$ and $22$ bits if $w=128$, will all be quantized to $0$, respectively. With the same quantized reliability values, even though sorting is performed, the SABM-SCC decoder cannot effectively find the most unreliable $3$ bits (out of the $5$ or $22$ bits with $|\lambda_{l,k}|_q=0$). This will make BF more likely to flip the wrong bits, and thus cause a performance loss.

The analysis and discussion above give an intuition on the performance of the iSABM-SCC and SABM-SCC decoders with the proposed quantization scheme. A precise comparison is presented in Sec. \ref{sec:results}.

\subsection{Numerical Results}\label{sec:results}

\begin{figure}[!tb]
\centering
\includegraphics[width=0.48\textwidth]{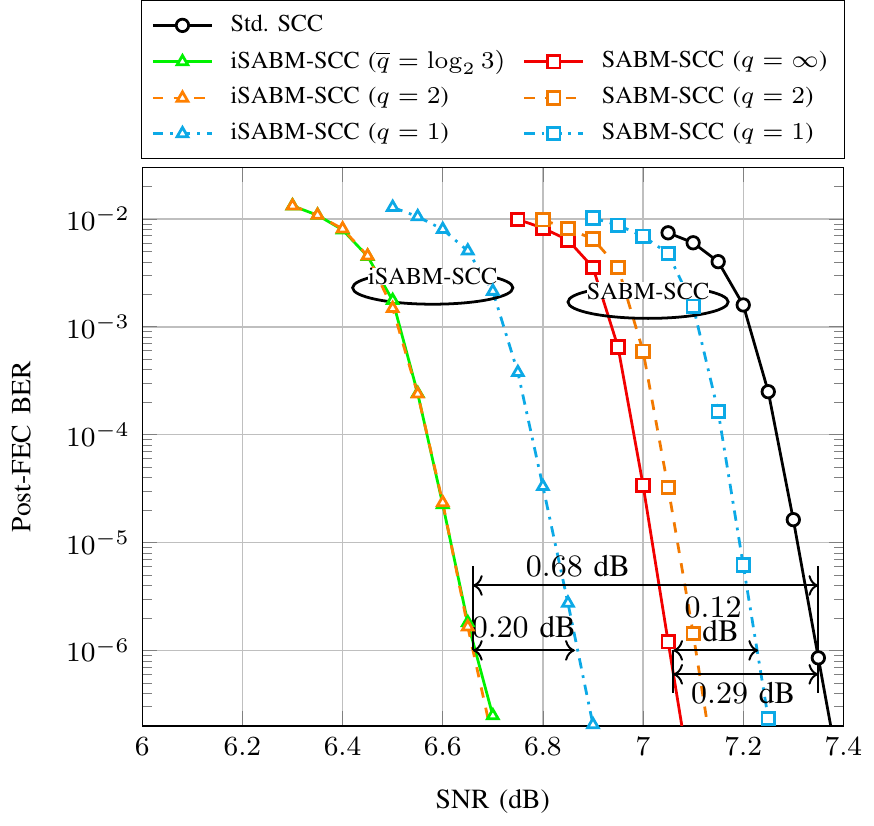}
\caption{Post-FEC BER vs. SNR for SABM- and iSABM-SCC decoding with 1-bit and 2-bit reliability quantization. The SCC code rate is $R=0.87$, and the modulation format is 2-PAM.}
\label{fig:PAM2-iSABM-Quantization}
\end{figure}

Fig.~\ref{fig:PAM2-iSABM-Quantization} shows the results of iSABM-SCC decoding with 2-bit (orange dashed curve) and 1-bit (blue dashed dotted curve) reliability quantization and {$L-K=7$}. The modulation format is 2-PAM. Two baselines are: standard SCC decoding and iSABM-SCC decoding with an idealized three-level quantizer (as shown in Fig.~\ref{subfig:woquantizationPDF}). In theory, the latter case requires $\overline{q}=\log_{2}3$ quantization bits, where  $\overline{q}$ is used to denote the non-integer quantization bits.
As predicted in the third to last paragraph of Sec. IV-A, iSABM-SCC with $2$-bit reliability quantization suffers negligible performance loss, while that with $1$-bit reliability quantization has a larger performance loss, i.e., $0.20$~dB. However, the latter can efficiently reduce the occupied memory for storing marked information by $50\%$.

Fig.~\ref{fig:PAM2-iSABM-Quantization} also shows the impact of reliability quantization on the performance of the SABM-SCC decoder. Differently from iSABM-SCC, the idealized case for SABM-SCC is that without quantization, or in other words, with $q=\infty$ quantization. As Fig.~\ref{fig:PAM2-iSABM-Quantization} shows, $1$-bit reliability quantization will degrade the performance by $0.12$~dB. Although the performance loss of SABM-SCC with 2-bit reliability quantization becomes smaller, there exists a clear gap to that without quantization. It indicates that SABM-SCC is more sensitive to the reliability quantization than iSABM-SCC. This is due to the fact that SABM-SCC needs the knowledge of the sorted most unreliable $d_0-t-1$ HUBs. However, as explained in Sec. IV-A, the same quantized value of $|\lambda_{l,k}|_q$ of the HUBs makes the decoder unaware of which HUB is the most unreliable one. In this case, the decoder always simply takes the required number of HUBs from left to right in each row of a SCC block for flipping.

\begin{figure}[!tb] 
\centering
\includegraphics[width=0.48\textwidth]{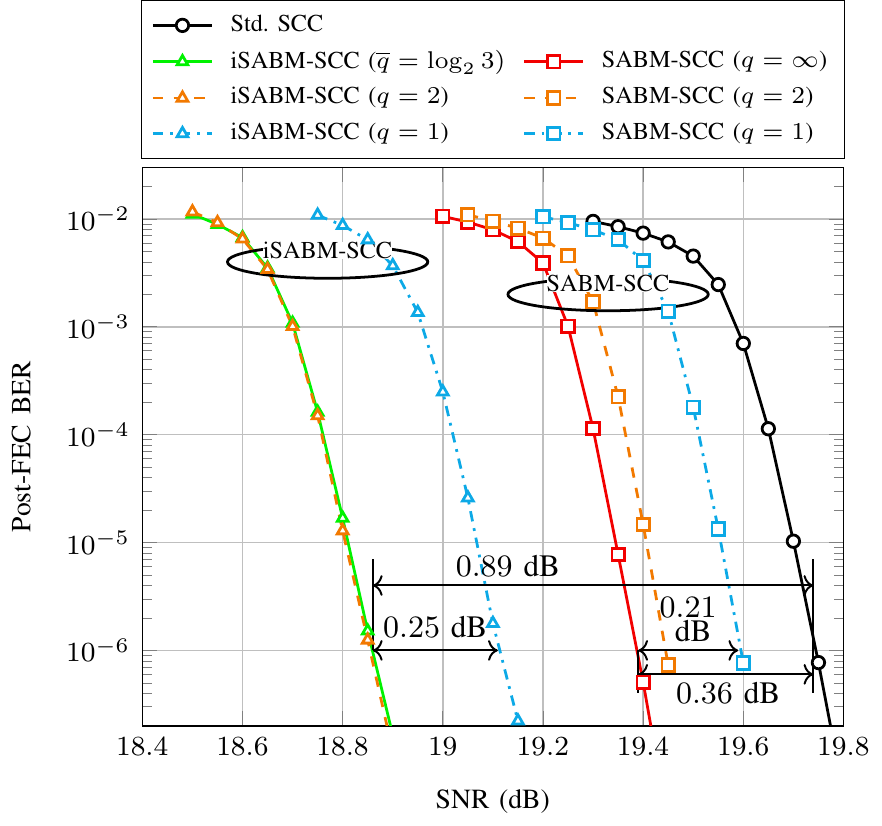}
\caption{Post-FEC BER vs. SNR for SABM- and iSABM-SCC decoding with 1-bit and 2-bit reliability quantizations. The SCC code rate is $R=0.87$, and the modulation format is 8-PAM.}
\label{fig:PAM8-iSABM-Quantization}
\end{figure}

Fig.~\ref{fig:PAM8-iSABM-Quantization} shows the influence of reliability quantization on the SABM- and iSABM-SCC decoders for 8-PAM. We can also find that 2-bit reliability quantization is enough to enable a negligible performance loss for iSABM-SCC decoding. To further relax the requirement on hardware, $1$-bit reliability quantization is also feasible, but with $0.25$~dB ($0.21$~dB) performance loss for the iSABM-SCC (SABM-SCC) decoders.

\section{Complexity Analysis}
The channel LLRs in SABM- and iSABM-SCC are only used to classify bits into HUBs, HRBs, and UBs. In general, the complexity of the SABM- and iSABM-SCC decoding is much lower than that of SD decoding (see e.g.,~\cite{Douxin_ISTC2018,SCC_LDPC2020}). Due to the fact that the SABM and iSABM algorithms need to have the HUB and HRB information, this overhead makes the SABM- and iSABM-SCC decoding more complicated than HD decoding (e.g.,~\cite{Christian1,Holzbaur2017}). Compared to SABM-SCC, the most important part of the proposed iSABM-SCC decoder is that the time-consuming sorting is completely avoided. Some implementation aspects of the SABM-SCC decoding have already been discussed in~\cite[Sec. IV]{YiTCOM2019} and~\cite[Sec. II]{Alex_OECC2019}. In what follows, we will mainly discuss the complexity increase of the newly proposed iSABM-SCC decoding, and compare with other two state-of-the-art SA-HD decoding methods: SABM-SR~\cite{Gabriele_ECOC2019} and BEE-SCC~\cite{AlirezaSCC2020}.

\subsection{Complexity Analysis}

With respect to standard SCCs, a big contribution to the complexity increase of the iSABM-SCC decoder is the extra BDDs coming from the second BDD attempt after BF (see Fig.~\ref{fig:iSABM} (right)). Let $\overline{N}$ and $N_{\text{sd}}$ be the number of BDDs in the iSABM-SCC decoding and standard SCC decoding within a window, respectively. The relative complexity increase caused by the additional BDDs is given by~\cite[eq.~(4)]{YiTCOM2019}
\begin{equation}
    \eta_1\triangleq\frac{\overline{N}-N_{\text{sd}}}{N_{\text{sd}}}=\frac{\overline{N}-w(L-1)\ell}{{w(L-1)\ell}}.
    \label{Eq:complexityIncrease}
\end{equation}

Considering for example a SCC with BCH $(256,239,2)$ component code, $N_{\text{sd}}$ equals to $7,168$ when $L=9$ and $\ell=7$. Since the number of BDDs within an iSABM-SCC decoding window is not deterministic, we calculated $\overline{N}$ as the average value of the first $10,000$ windows. iSABM-SCC with $K=2$ resulted in a $\eta_1=22.35\%$  relative complexity increase at an SNR of $6.57$~dB. It can be expected that this value will become smaller, as channel SNR increases. This is due to the reduced channel errors, which will result in less BDD failures and miscorrections in the first BDD attempt.

It should be mentioned that depending on how the algorithms in this paper are implemented, the complexity increase given by \eqref{Eq:complexityIncrease} could be an underestimation of the true complexity increase. The reason is as follows. BDD can be implemented as a syndrome-based decoder, which performs syndrome calculation first, and then estimates the error pattern using syndromes and corrects errors. We refer to this syndrome calculation, error estimation and correction as \emph{full BDD}. When the calculated syndrome vector is zero, BDD stops as its input is a valid codeword. We refer to this as \emph{partial BDD}, which is less complex than full BDD. The expression in \eqref{Eq:complexityIncrease} assume all decoding attempts are full BDD. However, standard SCC rarely needs full BDDs, as the code rate is often adapted to the channel condition. On the other hand, the extra BDDs in iSABM-SCC always need error pattern estimation, as their input is either a miscorrection or a failure (which corresponds to a nonzero syndrome). This makes the actual relative complexity increase potentially higher than that calculated from (\ref{Eq:complexityIncrease}).

Let $v_\text{sc}$ and $v_\text{ep}$ be the time required for syndrome calculation and error pattern estimation, respectively. For the $i$th window, a more accurate relative complexity increase (due to the extra BDDs) is
\begin{equation}
   \eta_2 \triangleq
   \frac{\sum\limits_{h=1}^{\ell}{\sum\limits_{p=1}^{L-1}(\overline{D}_{hp}v_\text{sc}+\overline{P}_{hp}v_\text{ep})}}{\sum\limits_{h=1}^{\ell}{\sum\limits_{p=1}^{L-1}(D^\text{sd}_{hp}v_\text{sc}+P^\text{sd}_{hp}v_\text{ep})}}-1,
    \label{Eq:complexityIncrease2}
\end{equation}
where $D^\text{sd}_{hp}$ and $\overline{D}_{hp}$ denote the number of syndrome calculations of component words in $[\boldsymbol{Y}^{T}_{i+p-1} \boldsymbol{Y}_{i+p}]$ at $h$th iteration of standard SCC and iSABM-SCC, respectively. In \eqref{Eq:complexityIncrease2}, $P^\text{sd}_{hp}$ and $\overline{P}_{hp}$ denote the number of error pattern estimations of standard SCC and iSABM-SCC, respectively. For standard SCC, $P^\text{sd}_{hp}\leq D^\text{sd}_{hp}\leq w$. Since iSABM is performed over the last $L-K$ SCC blocks, $\overline{P}_{hp}\leq \overline{D}_{hp}\leq2w$ for $p>K$ (due to the extra BDDs), and $\overline{P}_{hp}\leq\overline{D}_{hp}\leq w$ for $p\leq K$.

In (\ref{Eq:complexityIncrease2}), $v_\text{sc}$ and $v_\text{ep}$ (or, equivalently, the ratio of $v_\text{sc}$ to $v_\text{ep}$) are the key to estimate $\eta_2$ accurately. However, their exact values depend on many implementation-specific details such as, for example, the particular hardware architecture and the degree of parallelization. Therefore, an objective and accurate analysis for $v_\text{sc}$ and $v_\text{ep}$ as well as $\eta_2$ is beyond the scope of this paper and is left as future work.

Another important contribution to the complexity increase of the iSABM-SCC decoder is marking bits. For every HD-estimated bit $\hat{b}_{l,k}$, the iSABM-SCC decoder needs to mark whether it is an HUB, HRB, or UB (see~\eqref{BitMarking}). Bit marking occurs only once before decoding. During decoding, the marked information is not updated, which is statically stored in a data random access memory (RAM). From a hardware implementation point of view, this bit marking process is quite simple as only two comparator circuits with thresholds of $\delta_1$ and $\delta_2$ are required. For the storage of the marked information, {$2$ bits are enough to indicate the three possibilities: HRB, HUB, and UB. To be a more hardware-friendly FEC code, the simulation results shown in Sec.~IV-B have demonstrated that $1$-bit representation of a $|\lambda_{l,k}|$ value is also feasible at the expense of a small performance loss, but with $50\%$ memory save.}

In addition, the iSABM-SCC decoder also requires the syndrome information for each component codeword to perform miscorrection detection. However, this is costless, as BDD will naturally compute (and store) the syndromes for each component codeword.

To randomly determine the bit flippings at each iteration, the need for a random number generator (RNG) is also an overhead to the iSABM-SCC decoder. One of the most simple and common way to generate pseudo-random numbers is using linear-feedback shift register (LFSR). The potential problem is that the number of HUBs in each component codeword is not deterministic. If the worst case is considered, i.e., the $2w$ bits of a component codeword are all HUBs, the required number of bits for the LFSR is $\log_2(2w+1)$.
Therefore, the  accurate complexity of RNG depends on the practical hardware implementation, which is left for future work.

\subsection{Approximate Complexity Discussion}

Table~\ref{tab:complexity comparison} shows a brief complexity comparison between the iSABM-SCC, SABM-SR-SCC, and BEE-SCC decoders, which we will discuss below. They are three kinds of SA-HD decoders that have been proposed very recently, and present comparable performance (the performance comparison has been shown in Sec. III-B). However, we would like to highlight that the iSABM-SCC decoder proposed in this paper is the simplest one. The reasons for this can be explained from four aspects:
\begin{itemize}
   \item \textbf{Less BDD attempts}: As Fig.~\ref{fig:iSABM} shows, iSABM-SCC performs SA decoding over part of the SCC blocks within a window, i.e., $L-K$ out of $L$ SCC blocks. This corresponds to $K$ groups of BDDs and $L-K-1$ groups of iSABMs at each iteration. SABM-SR-SCC is similar to iSABM-SCC, but with $L-K-1$ groups of SABMs with \emph{scaled reliabilities}. By contrast, BEE-SCC performs $L-1$ groups of SA component decodings at each iteration. In terms of the SA decoding, the three algorithms all need to perform BDD multiple times to decode a component word. However, the number of BDD attempts in iSABM-SCC and SABM-SR-SCC is at most twice, while that in the BEE-SCC is three times (one is from the upper branch shown in~\cite[Fig. 8]{AlirezaSCC2020}, while the other two are from the bottom branch due to the erasure decoding). The accumulated large amount of extra BDDs in BEE-SCC will result in a higher complexity increase.

\begin{table}[!t]
\renewcommand{\arraystretch}{1.3}
\newcommand{\tabincell}[2]{
\begin{tabular}{@{}#1@{}}#2\end{tabular}}
\caption{Complexity comparison between iSABM-, SABM-SR-, and BEE-SCC decoders}
\centering
{
\resizebox{0.5\textwidth}{!}{
\begin{tabular}{c|c|c|c}
\hline

\hline
 & iSABM-SCC & SABM-SR-SCC~\cite{Gabriele_ECOC2019} & BEE-SCC~\cite{AlirezaSCC2020}  \\
\hline

\hline
\tabincell{c}{{Component}\\ {decoders}} & \multicolumn{2}{c|}{\tabincell{c}{$K$ groups of BDDs and  \\ $L-K-1$ groups of SA\\ component decodings}} & \tabincell{c}{$L-1$ groups of \\SA component \\decodings } \\
\hline
\tabincell{c}{BDDs in  a \\SA component \\ decoding} & \multicolumn{2}{c|}{ $1$ or $2$ (random)} & 3 \\
\hline
LLR sorting &  No & \tabincell{c}{Yes \\ (find the most \\  unreliable \\ $d_0-t-1$ \\bits)} & \tabincell{c}{Yes \\ (find the  most \\ unreliable $2$\\ bits)}  \\
\hline
LLR updating & No & \multicolumn{2}{c}{ Yes} \\
\hline
\tabincell{c}{Extra memory\\ required} & \tabincell{c}{ Marked info. }  & \tabincell{c}{Channel LLRs, \\ and marked info. }  & \tabincell{c}{Channel LLRs, \\ LUTs, and\\ ternary messages \\  for LLR update} \\
\hline
\tabincell{c}{Main contribution\\ to the extra  \\data-flow w.r.t. \\ standard SCCs}    &  \tabincell{c}{Read marked info. \\from data RAM}  & \multicolumn{2}{c}{\tabincell{c}{Read channel LLRs from data RAM, \\and update LLRs}} \\
\hline

\hline
\end{tabular}
}}
\label{tab:complexity comparison}
\vspace{-4ex}
\end{table}

    \item \textbf{No LLR sorting and updating}:
    Both SABM-SR-SCC and BEE-SCC update their LLRs as a function of the iterations. Using the updated LLRs, SABM-SR-SCC updates the HRBs by comparing with a reliability threshold. However, the HUBs are updated by performing reliability sorting to determine the sorted $d_0-t-1$ HUBs in each component codeword. In the BEE-SCC decoder, reliability sorting is also required to find the most unreliable $2$ bits to determine the $2$ erasures in each component codeword. It is well known that sorting process will greatly increase the complexity of the algorithms. In contrast, the iSABM-SCC decoder determines the HRBs and HUBs by simply comparing the absolute LLR values with two reliability thresholds. The marked reliability information are not updated either.

    \item \textbf{Reduced memory usage}: In terms of the required extra memories (with respect to standard SCC), {the iSABM-SCC decoder needs to store $2$-bit marked information for each bit $\hat{b}_{l,k}$ (rather than the LLRs)}. By contrast, the SABM-SR-SCC decoder also requires to store the channel LLRs with multiple bits, and so does BEE-SCC. In addition to the channel LLRs, BEE-SCC also requires memory for storing look-up tables (LUTs) and ternary messages (from the output of BDD and error-and-erasure decoding (EED) in~\cite[Fig. 8]{AlirezaSCC2020}) for LLR update. Although the storage of the LUTs occupies little additional memory, the mathematical computation of the elements in the LUTs also needs to be considered~\cite[Eq. (13)]{AlirezaSCC2020}).
    \item \textbf{Reduced data-flow}:
   In terms of data-flow between the component decoders, iSABM-SCC has a similar complexity as SABM-SR-SCC and BEE-SCC, i.e., only binary bits are exchanged. However, iSABM-SCC has a reduced data-flow between the component decoder and data RAM. The main contribution to the extra data-flow (with respect to standard SCCs) is reading the marked information for each bit $\hat{b}_{l,k}$ from a statically stored bit-marking database. By contrast, SABM-SR-SCC requires to read the channel LLRs and the weight factors from the data RAM first. After updating LLRs and remarking bits, the HRB and HUB information can then be delivered to the decoder to perform SABM decoding. Similarly, BEE-SCC needs to read the channel LLRs, LUTs, and ternary messages (from the output of BDD and EED) for LLR update.
\end{itemize}

\section{Conclusions}
In this paper, a new soft-aided hard-decision decoder, called iSABM-SCC, is proposed to improve the decoding of staircase codes. The iSABM-SCC decoder is based on modifications of the soft-aided bit-marking algorithm with a simplified bit marking process and an improved bit flipping strategy. By performing soft-aided decoding over multiple SCC blocks, gains up to $0.53$~dB with respect to SABM-SCCs and up to $0.91$~dB with respect to standard SCCs are reported. The analysis of reliability quantization on the performance of iSABM-SCC decoder show that $2$-bit representation will cause negligible performance loss, { while $1$-bit representation causes a $0.25$~dB gain penalty but with $50\%$ memory saving for the storage of marked information}. The retained gains are still much higher than those provided by the SABM-SCC decoder with infinite-bit reliability quantization. Due to the excellent performance and hardware-friendly implementation, we believe iSABM-SCC will be a very appealing FEC solution for future high-speed low-cost optical fiber communication systems, especially now that SCCs have been recommended in the standards for 100G-LR and 400G-ZR optical transport networks.

\bibliographystyle{IEEEtran}
\bibliography{refs}

\end{document}